# Imaging GHz surface acoustic wave modes in electrostricted $LaAlO_3/SrTiO_3$ heterostructures

Ranjani Ramachandran ; Sayanwita Biswas ; Prithwijit Mandal ; Kyoungjun Lee; Madeleine Msall ; Chang-Beom Eom ; Patrick Irvin ; Jeremy Levy ; Mingyun Yuan

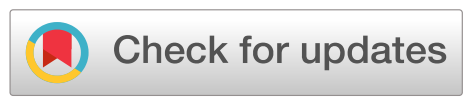

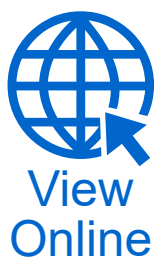
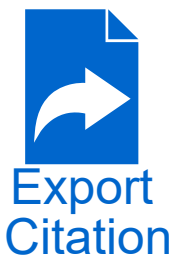

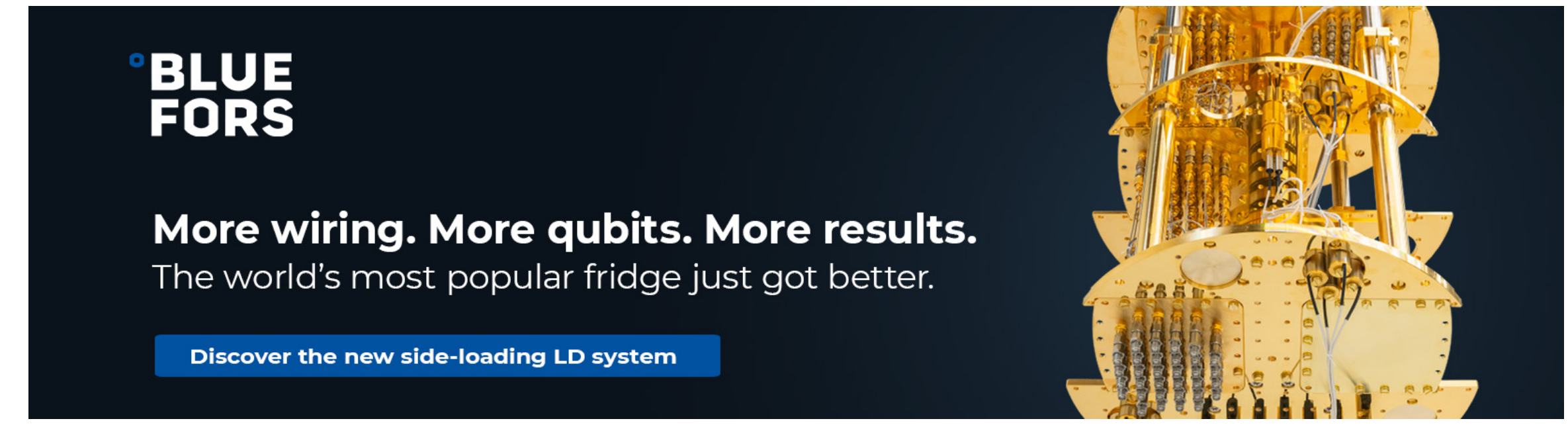

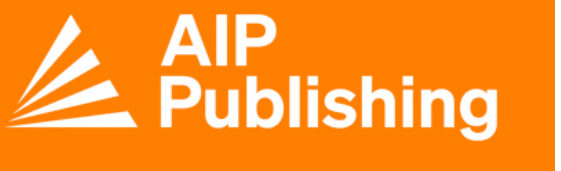

# Imaging GHz surface acoustic wave modes in electrostricted $LaAlO_3/SrTiO_3$ heterostructures

View Online

Export Citation

CrossMark

Ranjani Ramachandran,[1,2,3] Sayanwita Biswas,[1,2] Prithwijit Mandal,[4] Kyoungjun Lee,[4] Madeleine Msall,[5] Chang-Beom Eom,[4] Patrick Irvin,[1,2] Jeremy Levy,[1,2] and Mingyun Yuan[3,a)]

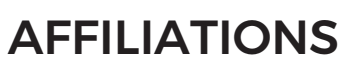
AFFILIATIONS

[1]Department of Physics and Astronomy, University of Pittsburgh, Pittsburgh, Pennsylvania 15260, USA
[2]Pittsburgh Quantum Institute, Pittsburgh, Pennsylvania 15260, USA
[3]Paul-Drude-Institut für Festkörperelektronik, Leibniz-Institut im Forschungsverbund Berlin e.V., Berlin 10117, Germany
[4]Department of Materials Science and Engineering, University of Wisconsin-Madison, Madison, Wisconsin 53706, USA
[5]Bowdoin College, Brunswick, Maine 04011, USA

[a)]**Author to whom correspondence should be addressed:** yuan@pdi-berlin.de

## ABSTRACT

The $LaAlO_3/SrTiO_3$ (LAO/STO) interface hosts a gate-tunable superconducting two-dimensional electron gas (2DEG), which can be programmed to create quantum devices, such as ballistic electron waveguides and quantum dots. To fully exploit this platform for quantum transport, a key requirement is the ability to shuttle single electrons, electron pairs, and other exotic states between spatially separated devices with precision. Surface acoustic waves (SAWs), which travel along the surface of a solid, offer a powerful route to achieve this through their moving electrical potential that captures and transfers electrons. In particular, SAWs in the GHz regime enable fast, controlled transport of individual quantum particles. Although this approach is well-explored in GaAs-based 2DEG, SAW generation in STO remains largely unexplored due to the lack of intrinsic piezoelectricity at room temperature. Here, we investigate room-temperature SAWs in LAO/STO and observe SAW modes up to 2.2 GHz with very low propagation loss of the order $10^{-3}$ dB per wavelength. To directly visualize these modes, we employ atomic acoustic force microscopy, achieving sub-micron resolution imaging of the SAW wave forms, providing insight into the electrostriction-induced SAW generation mechanism. Our measurements indicate a shear horizontal-type mode, which provides the ability to couple to in-plane degrees of freedom for future acoustoelectric and quantum device applications. This work studies the fundamentals of SAW excitation and propagation on STO, a widely used and commercially available substrate, enabling straightforward coupling of SAWs to a broad range of materials that can be grown or transferred onto STO.

The $LaAlO_3/SrTiO_3$ (LAO/STO) interface hosts a gate-tunable high-mobility 2-dimensional electron gas (2DEG),[1,2] which becomes superconducting at low temperature.[3] Quantum devices such as single-electron transistors[4–6] and ballistic electron waveguides[7] have been demonstrated at this interface, making it a promising platform to explore novel quantum applications. A central challenge for such quantum systems is the ability to shuttle single electrons, electron pairs,[8] and other exotic states[9] between spatially separated devices in a reproducible and precise fashion.

Surface acoustic waves (SAWs), which are periodic elastic deformations that propagate along the surface of a solid, offer a dynamic route to achieve this. In piezoelectric materials, the SAW strain deformation is accompanied by an electric potential wave that can couple efficiently to mobile charges and drive acoustoelectric currents.[10–12] SAWs have been coupled to various quantum devices to realize the long-range transfer of electrons,[13] excitons,[14] and spins.[15] 2DEGs that can be combined with SAWs that have low propagation loss at GHz frequencies are attractive platforms for fast control of electrons and acoustoelectric effects. For example, (Al,Ga)As/GaAs heterostructures,[16,17] with intrinsic piezoelectricity in combination with high-quality quantum transport channels and high-frequency operation enable precise, fast electron shuttling for quantum devices.[13,18–21]

SAW generation in LAO/STO 2DEG systems is hindered by the fact that bulk $SrTiO_3$ (STO) is centrosymmetric at room temperature [Fig. 1(a)] and lacks piezoelectricity.[22] A DC electric field can, nevertheless, distort the lattice into a tetragonal structure [Fig. 1(b)], inducing a net polarization through electrostriction, enabling SAW generation.[12,23–26] Despite these demonstrations, this method for SAW generation remains relatively underexplored. Of particular interest are the effects of structural phase transition of STO, such as the antiferrodistortive transition at 110 K[27,28] and paraelectric quantum phases below 10 K[29] on SAWs. Some experiments report that STO SAWs disappear below 105 K,[12,30] while others report that they persist to lower temperatures.[24,31] It is also unknown whether electrostricted-SAWs are spatially uniform like their piezoelectric counterpart, and modeling of the effect of the net-polarization that is induced locally near the IDTs remains unchartered. Better understanding of the full range of STO SAW behavior will enhance its applicability to the study of important physical systems and provide a versatile SAW platform that can readily be integrated with materials such as graphene[32,33] and transition-metal dichalcogenides (TMDs),[34,35] to explore acoustoelectronics, electron–phonon and phonon–magnon interactions,[36–39] sensing, and liquid-loading applications.[40–42]

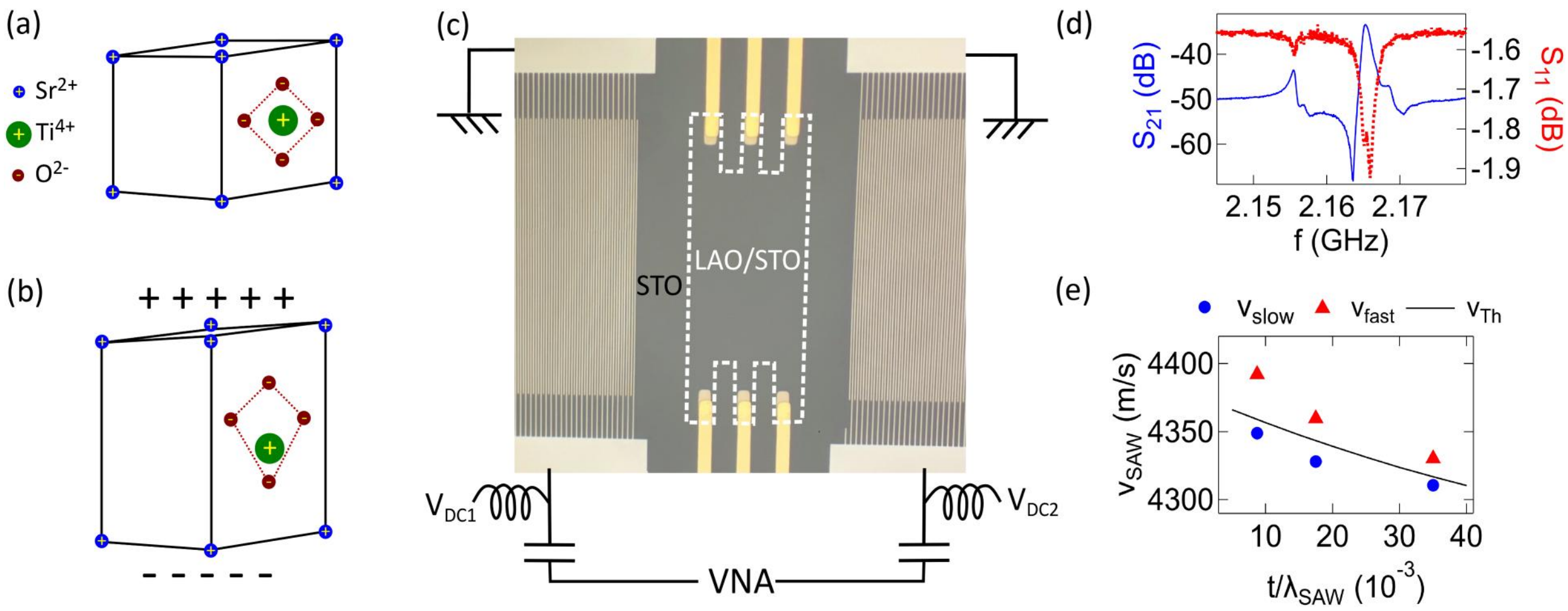

**FIG. 1.** (a) Illustration of centrosymmetric STO in the absence of electric field. (b) Illustration of a tetragonal structural deformation when a DC electric field is applied on STO. (c) Schematic of the experimental setup used to measure the S-parameters. A bias-tee is connected between the IDTs and the VNA to apply the DC bias $V_{DC1}$ and $V_{DC2}$. (d) S parameters measured from delay line DL2 of $\lambda_{SAW} = 2\,\mu$m with $V_{DC1} = V_{DC2} = 30$ V. (e) Velocity of SAW modes as a function of normalized thickness $(t/\lambda_{SAW})$ under $V_{DC1} = V_{DC2} = 30$ V (solid symbol) against the calculated curve, $v_{Th}$.

In this work, we investigate electrostriction-mediated SAW modes on LAO/STO (001) at room temperature over a frequency range from 500 MHz to 2.2 GHz. Operating at previously underexplored GHz frequencies and shortened wavelengths $\lambda_{SAW}$ is vital for hybrid quantum device applications. Short wavelength SAWs offer stronger potential confinement over smaller length scales, enabling single-electron trapping and faster triggering rates for quantum devices. More importantly, GHz operation can access energy scales that allow coupling to additional degrees of freedom such as spin, subband separation or magnon excitations. Our measurements reveal two distinct SAW modes separated in frequency by 10 MHz (near 2.16 GHz) and record ultra-low propagation loss ($\gamma$) as small as $6\times\ 10^{-3}$ dB/$\lambda_{SAW}$. We also image the acoustic field distribution with nanometer-scale resolution using atomic acoustic force microscopy (AAFM), a powerful scanning-probe method for studying SAWs[43,44] and hybrid nanomechanical systems in general.[45,46] The high-resolution spatial map of the acoustic response complements the frequency domain measurements of the scattering (S)-parameters of the delay line formed by a pair of IDTs and unambiguously confirms the surface proximity of the acoustic wave propagation. Based on the AAFM, we attribute the lower frequency to the expected Rayleigh-SAW mode and the higher frequency mode to a shear horizontal (SH)-type SAW mode. These two polarizations would otherwise be indistinguishable solely with the S-parameter measurement. In addition, this nanoscale imaging can reveal undesired scattering near metal electrodes or at the LAO/STO boundary, which can strongly impact SAW coupling to the 2DEG. It also clarifies whether there are local variations in electrostrictive SAW generation, providing valuable insight for new SAW device designs on STO and other non-conventional materials.

Figure 1(c) shows an illustration of a typical sample layout and the experimental setup. We started with four unit cells of LAO on STO (001) grown as described in the supplementary material, Sec. I and selectively etched away the LAO by reactive ion etching[47] to define the “canvases” containing the 2DEG, shown by the white dotted line. The IDTs were patterned using standard electron beam lithography and metalized with a stack of Ti/Al/Ti with thicknesses of 10/50/10 nm, leading to a metallization thickness of $t = 70$ nm. DC biases, $V_{DC1}$ and $V_{DC2}$, are applied across IDT fingers of opposite polarities in the input and output IDTs, respectively, to induce electrostriction on STO. We measured five IDT delay lines (DL) with wavelengths $\lambda_{SAW}$ of 2, 4, and 8 $\mu$m (Table I). All IDTs have an aperture of 100 $\mu$m and 50 (DL8) or 100 (all others) finger pairs. All measurements were performed at room temperature and ambient pressure, with a nominal RF power of $P_{RF} = 0$ dBm unless otherwise specified.

We characterize the RF response of the delay lines in the frequency domain using a vector network analyzer (VNA). Figure 1(d) shows the S-parameters for DL2, with $\lambda_{SAW} = 2\,\mu$m under $V_{DC1} = V_{DC2} = 30$ V. We observe two distinct sets of transmission peaks, indicating two SAW modes with characteristic frequencies $f_{slow} = 2.156$ GHz and $f_{fast} = 2.165$ GHz. The S-parameters of other devices are included in the supplementary material, Sec. IV. The fast mode consistently exhibits a stronger dip in the reflection coefficient ($S_{11}$) and peak in the transmission coefficient ($S_{21}$) than the slow mode across all measured devices. Interestingly, this two-mode behavior becomes more distinct with higher frequency [Figs. S4(a)–S4(c)], which could explain why previous report[12] probing a lower frequency of $f_{SAW}$ = 445 MHz ($\lambda_{SAW}$ = 10 $\mu$m) recorded only a single SAW mode in STO.

**TABLE I.** List of devices measured in this work along with their wavelength and measured propagation loss. DL stands for the delay line, and the number following DL is the wavelength.

| Device | $\lambda_{SAW}$ ($\mu$m) | $\gamma$ (dB/$\mu$m) | $\gamma$ (dB/$\lambda_{SAW}$) |
|---|---|---|---|
| DL8 | 8 | ... | ... |
| DL4a | 4 | 0.001 80 ± 0.000 91 | 0.007 20 ± 0.003 64 |
| DL4b | 4 | 0.001 47 ± 0.000 61 | 0.005 90 ± 0.002 43 |
| DL4c | 4 | 0.001 58 ± 0.000 13 | 0.006 32 ± 0.000 51 |
| DL2 | 2 | 0.006 05 ± 0.000 19 | 0.012 10 ± 0.000 39 |

To gain better understanding of the SAW behavior, we compare the experimental SAW velocity (solid symbols) as a function of normalized thickness (the ratio of metallization thickness by IDT wavelength, $t/\lambda_{SAW}$), to the expected Rayleigh mode velocity (black line) in Fig. 1(e). Both modes are in close proximity to the calculation, establishing that both are surface acoustic modes. The slight dispersion of $v_{SAW}$ with the normalized thickness originates from the loading of the metalized thin film forming the IDTs. The STO is treated as a bulk substrate, and its elastic constants are taken from Refs. 48 and 49. More details of the numerical calculation of the expected Rayleigh mode velocity, $v_{Th}$, can be found in the supplementary material, Sec. II.

Figure 2 illustrates the effect of the varying DC bias that enables electrostrictive SAW generation. Figure 2(a) shows $S_{21}$ for a $\lambda_{SAW}$ = 4 $\mu$m device, DL4c, with a constant bias of $V_{DC1}$ = 30 V on the input IDT, while the bias on the output IDT, $V_{DC2}$, is varied. We only observe the SAW resonances when both $V_{DC1}$ and $V_{DC2}$ are nonzero. Notably, both modes at $f_{slow}$ = 1.083 GHz (black arrow) and $f_{fast}$ = 1.089 GHz are enhanced with increasing $V_{DC2}$. The fast mode exhibits multiple oscillations as a result of Fabry–Pérot modes in the cavity formed by the opposing IDTs. The most prominent peak value of $S_{21}$ for the fast mode (near 1.089 GHz) shown in Fig. 2(a) is displayed in Fig. 2(b). This figure highlights the enhancement in transmission as $V_{DC2}$ increases and is consistent with previous reports.[12] The full curves can be found in Fig. S5.

Electrostriction-induced SAWs in STO are reported to show a memory effect, where the SAW signal is not immediately lost after $V_{DC}$ is turned off. This memory effect has been attributed to photoconductivity and oxygen vacancy migration.[50] We observe this effect in Fig. 2(c), which shows the decay of the $S_{21}$ maximum ($S_{21max}$) as a function of time $t$ after both bias voltages are set to zero. We fit the data to an exponential decay up to 3.5 min, plotted as red line for the whole vertical range, from which we extract a time constant of $\tau$ = 55.88 s. Remnants of the transmission resonance, however, persist for times beyond 23 min (supplementary material, Sec. III).

An important practical consideration when employing STO for SAW devices is the propagation loss, $\gamma$, which determines how efficiently acoustic energy is transmitted as it propagates away from the IDT. We measure $\gamma$ by probing the SAW echoes in the time domain [cf. Fig. 2(d)]. Here, the corresponding frequency range includes both the fast and the slow modes. Thus, it represents the general, rather than the mode-specific $\gamma$ for STO. The linear loss rates in Table I are extracted from a fit of the peak echoes. Our very low propagation losses (down to $\sim 10^{-3}$ dB/$\lambda_{SAW}$) are a macroscopic manifestation of the high quality of the surface and the layer interface, ensuring that scattering due to roughness and defects remains negligible, preserving the signal over millimeters. Remarkably, in the device with the highest recorded signal level, the SAW resonance enters a nonlinear regime and exhibits characteristics of a Duffing oscillator, as described in the supplementary material, Sec. IV, Fig. S4(g).

Next, we demonstrate the high-resolution spatial mapping of the SAW modes in DL2. Figure 3(a) shows the sample topography, including the nominally 0.5 $\mu$m wide fingers of the IDT on the left side of the image, measured by the DC tip deflection signal of the AAFM. A mechanical diode effect[44] associated with the quadratic response term of the tip-surface interaction enables the detection of the standing SAWs using a standard AFM cantilever. Since the GHz SAW frequencies are much higher than the operational mechanical frequencies of the cantilever, the RF excitation frequency applied to the generating IDT is amplitude-modulated by a lower-frequency sinusoid near one of the out-of-plane cantilever contact resonances at $f_{env} = 293$ kHz. Figure 3(b) shows $\delta_{SAW}$, the AC component of the tip deflection at $f_{env}$, which is proportional to the square of the SAW amplitude.[45,51] It is measured simultaneously with the topographic DC deflection, using standard lock-in detection techniques. The SAW oscillations are clearly visible, with a periodicity of 1 $\mu$m, as expected for standing waves formed within the delay line cavity with $\lambda_{SAW} = 2\,\mu$m. A further scan near the interface between the STO region and the LAO/STO region [Fig. 3(c)] shows that the SAWs are effectively transmitted through the interface, with negligible loss or scattering despite the significantly increased peak-to-valley roughness of 5 nm at the etch step (Fig. S2), which is ideal for coupling to the 2DEG.

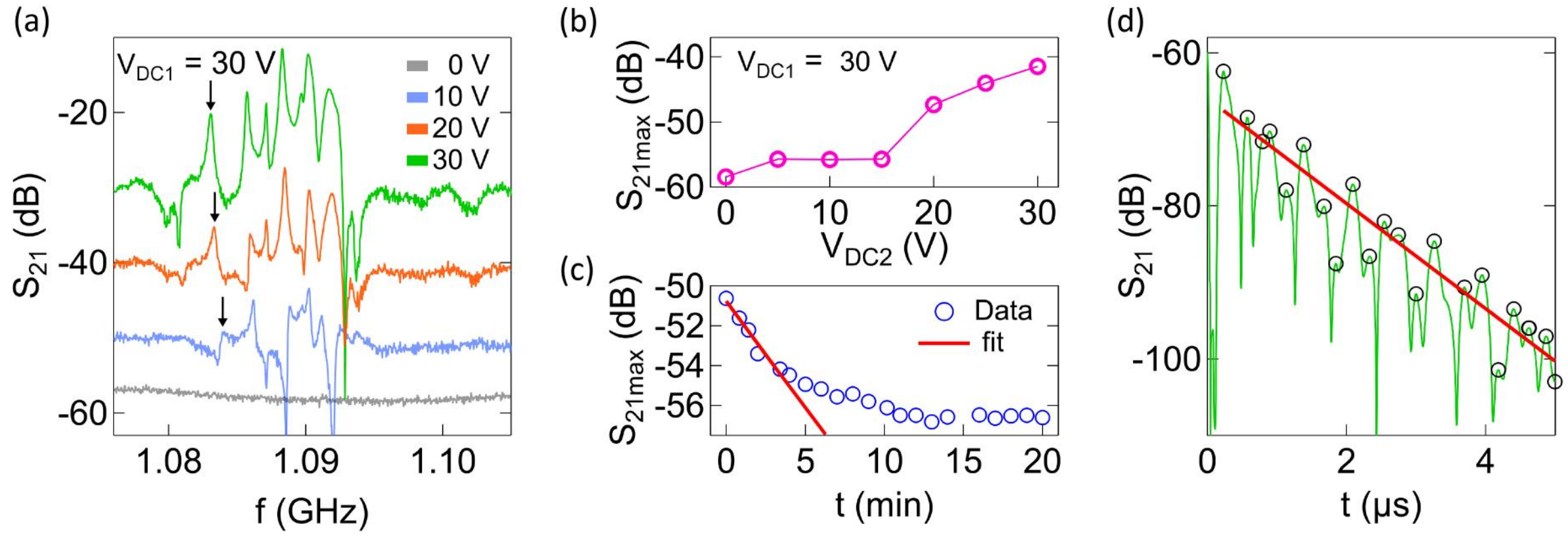

**FIG. 2.** (a) $S_{21}$ from DL4c with $\lambda_{SAW} = 4$ $\mu$m and $V_{DC1} = 30$ V and different values of $V_{DC2}$, with a 10 dB offset between adjacent curves. Black arrow indicates the $f_{slow}$ mode. (b) $S_{21max}$ as a function of $V_{DC2}$. (c) $S_{21max}$ decaying with time after zeroing $V_{DC1}$ and $V_{DC2}$ from 30 V. The initial response is fitted to an exponential decay $S_{21max} = Ae^{-t/\tau}$ with $A = -50.763$ dBm and $\tau = 55.88$ s. (d) Time-domain echoes of $S_{21}$ from DL4c with 250 $\lambda_{SAW}$ delay at $V_{DC1} = V_{DC2} = 30$ V. Peak values are fitted to a straight line to extract $\gamma$ listed in Table I.

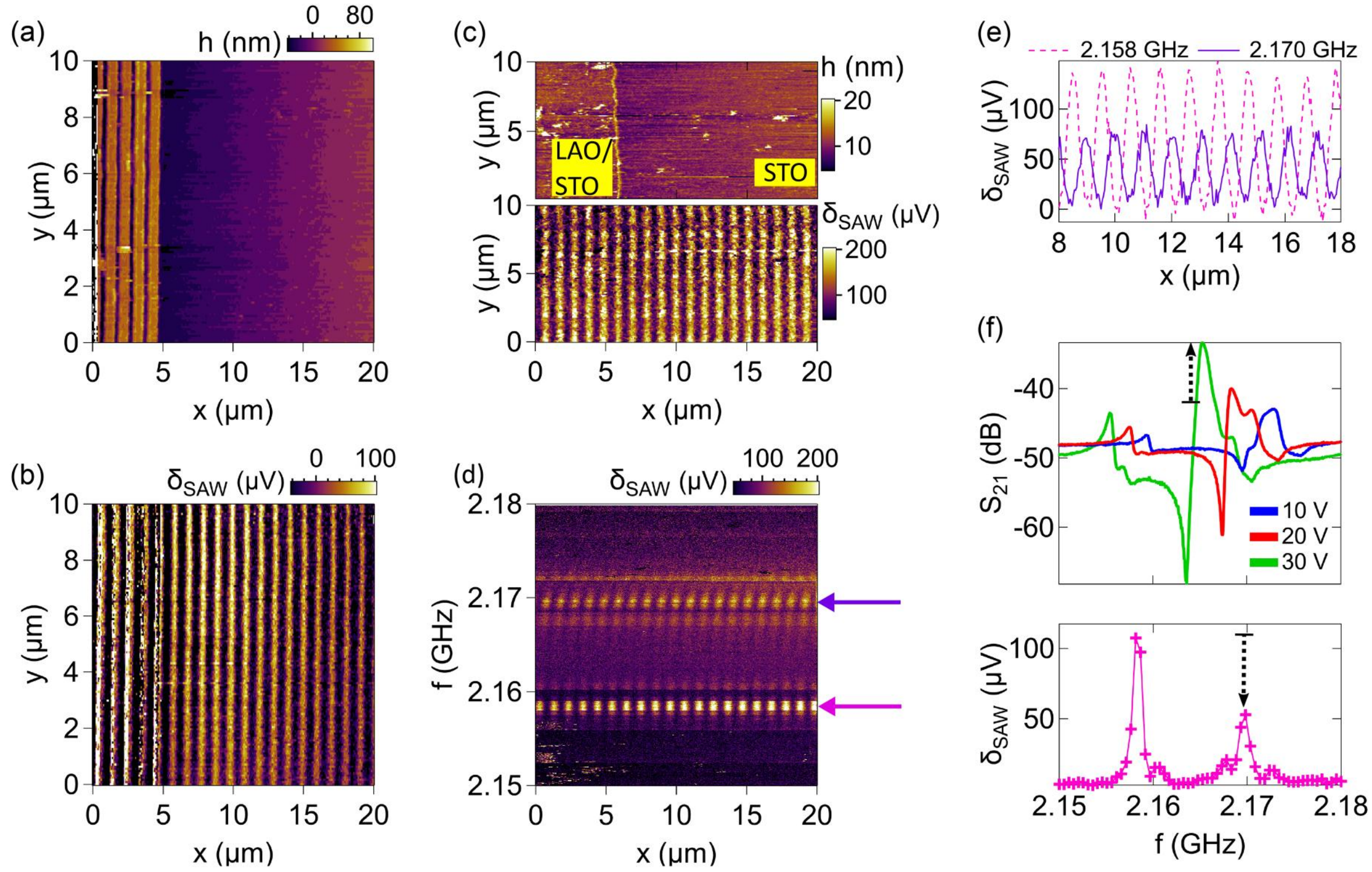

**FIG. 3.** AFM mapping of DL2 showing (a) sample topography and (b) the corresponding SAW profile $\delta_{SAW}$ corresponding to $P_{RF} = 10$ dBm, $f_{SAW} = 2.158$ GHz, and $f_{env} = 293$ kHz. Biases are kept at $V_{DC1} = V_{DC2} = 15$ V, and a protective series resistors were added to the DC sources. (c) Similar scans for a location where LAO is etched away in part of the sample. SAW propagation is not hindered by the step. (d) Scan $\delta_{SAW}$ in the *x*-direction as a function of applied RF frequency, *f*. (e) Cross section of the two SAW modes indicated by arrows in (d). (f) Comparison between the $S_{21}$ and peak value of $\delta_{SAW}$ averaged over $x = 20\,\mu$m for *f* stepping in (d). Arrows signal the opposite trends of the peak height.

Figure 3(d) shows AAFM measurements of the SAW signal along the delay line at different frequencies. Surprisingly, the SAW amplitudes are larger for the slower mode at 2.158 GHz than the faster mode at 2.17 GHz. In contrast, $S_{21}$ is consistently higher for the higher frequency mode, cf. Fig. 3(f). This difference is attributed to the sensitivity of our chosen cantilever mode to the out-of-plane (vertical) particle displacement. We hypothesize that the slow mode is a Rayleigh mode with coupled longitudinal and shear vertical (SV) displacements that are more effectively sensed by the cantilever. The fast mode could have predominantly shear horizontal (SH) displacements, which the cantilever is less sensitive to. Such SH-type SAWs are not expected in this frequency range from the numerical calculation based on centrosymmetric STO bulk and are not typically seen in an isotropic substrate. Surface effects[52] can be omitted due to the large thickness (1 mm) of the STO substrate. Our experimental observation suggests that for electrostrictive SAWs, it is no longer sufficient to consider STO as an isotropic material, and a structure of lower symmetry might be needed to fully capture the effective piezoelectric tensors.

This symmetry breaking can be attributed to the DC bias applied between the alternate IDT fingers. The electric fields from $V_{DC}$ induce electrostriction for the top few micrometers of the substrate, while the remainder of the bulk STO retains its centrosymmetric structure. This creates a waveguiding effect, similar to a layered structure comprising a piezoelectric thin film and a substrate, where additional SH-type SAWs are known to exist.[53] Rendering support to this hypothesis, in Fig. 3(f), the SAW resonance is pulled down as $V_{DC}$, and therefore, the effective depth of the electrostricted layer is increased, indicating a dispersion in velocity. This picture is in agreement with the model of a layered structure in which the sound velocity in the thin film is slower than in the substrate. Similar behavior is also reported in STO near the antiferrodistortive transition temperature, where the crystal structure changes from cubic to tetragonal.[54] Future experiments involving torsional force AFM imaging of the SAW modes can shed more light on the SH components.

In conclusion, we have demonstrated dual GHz electrostriction-mediated SAW modes with distinct displacement polarizations on LAO/STO. The measured low propagation loss on STO makes it a promising candidate for long-range transfer of qubits and practical integration with other materials that can be transferred onto STO. Access to multiple modes with different polarizations could enable selective probing of anisotropic strain-coupling. Our nanoscale AAFM imaging unambiguously confirms that both modes are SAWs, excluding explanations involving bulk acoustic modes that could be inadvertently generated in the substrate. Furthermore, it also verifies that unwanted scattering of SAWs and local variations in electrostriction are not a major concern for relevant future applications, validating the efficiency of electrostriction as a mechanism for SAW generation comparable to conventional piezoelectricity. The second, possibly SH-type, SAW mode was not identified before. Such SH-type modes have advantages over Rayleigh SAWs in effectively coupling to in-plane degrees of freedom for acousto-spin control[55,56] and sensing. STO is a widely used lattice-matched substrate for superconductors[57–60] (e.g., $YBa_2Cu_3O_{7-x}$ and FeSe) and magnetic oxides[61,62] (e.g., $La_{1-x}Sr_xMnO_3$ and $BiFeO_3$), opening access to acoustoelectric and acoustomagnetic phenomena in a broader range of materials compounds.

See the supplementary material for additional figures and experimental details.

We thank Yukihiko Takagaki and Alberto Hernández-Mínguez for a valuable discussion and Paulo V. Santos for advice and the SAW simulation framework. We thank Nazim Ashurbekov and Walid Anders for supporting the sample processing and experiment. R.R. and M.Y. acknowledge financial support from DAAD RISE Professional scholarship (No. 91954920). J.L. and C.-B.E. acknowledge support from ONR MURI under Grant No. N00014-21-1-2537. C.-B.E. acknowledges support for this research through a Vannevar Bush Faculty Fellowship (ONR N00014-20-1-2844), the Gordon and Betty Moore Foundation's EPiQS Initiative, Grant GBMF9065, and the National Science Foundation under Grant No. DMREF-2522669.

## AUTHOR DECLARATIONS

### Conflict of Interest

The authors have no conflicts to disclose.

### Author Contributions

**Ranjani Ramachandran:** Data curation (lead); Formal analysis (equal); Investigation (lead); Methodology (equal); Validation (equal); Visualization (lead); Writing – original draft (lead); Writing – review & editing (equal). **Sayanwita Biswas:** Investigation (supporting); Methodology (supporting). **Prithwijit Mandal:** Investigation (supporting); Methodology (supporting). **Kyoungjun Lee:** Investigation (supporting); Methodology (supporting). **Madeleine Msall:** Investigation (supporting); Methodology (supporting); Supervision (supporting); Writing – review & editing (supporting). **Chang-Beom Eom:** Funding acquisition (equal); Investigation (supporting); Methodology (supporting). **Patrick Irvin:** Investigation (supporting); Methodology (supporting). **Jeremy Levy:** Conceptualization (lead); Formal analysis (equal); Funding acquisition (equal); Investigation (supporting); Methodology (equal); Project administration (lead); Resources (lead); Supervision (equal); Writing – original draft (equal); Writing – review & editing (equal). **Mingyun Yuan:** Conceptualization (equal); Formal analysis (equal); Funding acquisition (equal); Investigation (equal); Methodology (lead); Project administration (equal); Supervision (lead); Validation (equal); Writing – original draft (equal); Writing – review & editing (lead).

## DATA AVAILABILITY

The data that support the findings of this study are available from the corresponding author upon reasonable request.